\documentclass[11pt,a4paper]{article}
\usepackage{jheppub}
\usepackage{slashed}
\usepackage{amsmath}
\usepackage{amssymb}
\usepackage{epsfig}

\def\({\left(} 
\def\){\right)}
\def\[{\left[} 
\def\]{\right]}
\newcommand{\non}{\nonumber \\}

\newcommand{\ie}{{\it i.e.,}\ }
\newcommand{\eg}{{\it e.g.,}\ }

\newcommand{\be}{\begin{equation}}
\newcommand{\ee}{\end{equation}}
\newcommand{\bea}{\begin{eqnarray}}
\newcommand{\eea}{\end{eqnarray}}

\newcommand{\mt}[1]{\textrm{\tiny #1}}

\catcode`\@=12

\def\del          {\partial}

\newcommand{\reef}[1]{(\ref{#1})}
\renewcommand{\eqref}[1]{(\ref{#1})}

\def\ph1{\phantom{1}}

\newcommand{\beq}{\begin{equation}}
\newcommand{\eeq}{\end{equation}}
\newcommand{\ba}{\begin{aligned}}
\newcommand{\ea}{\end{aligned}}
\newcommand{\beqa}{\begin{eqnarray}}
\newcommand{\eeqa}{\end{eqnarray}}
\newcommand{\beqar}{\begin{eqnarray*}}
\newcommand{\eeqar}{\end{eqnarray*}}

\title{On quantum quenches at one loop}

\author{Mikhail Goykhman}
\emailAdd{michael.goykhman@mail.huji.ac.il}

\author{Tom Shachar}
\emailAdd{tom.shachar@mail.huji.ac.il}

\author{and Michael Smolkin}
\emailAdd{michael.smolkin@mail.huji.ac.il}
\affiliation{The Racah Institute of Physics, The Hebrew University of Jerusalem, \\ Jerusalem 91904, Israel}

\abstract{ We study global quenches in a number of interacting quantum field theory models away from the conformal regime.
We conduct a perturbative renormalization at one-loop level and track the modifications of the quench protocol induced by the renormalization group flow. The scaling of various observables at early times is evaluated in the regime of rapid quench rates, with a particular emphasis placed on the leading order effects that cannot be recovered using the finite order conformal perturbation theory. We employ the canonical ideas of effective action to verify our results and discuss a potential route towards understanding the late time dynamics.
}

\begin{document}

\maketitle

\section{Introduction}

The path of getting from the Schrodinger equation at the fundamental level to the emergence of statistical mechanics at the macroscopic scale is far from being understood in a satisfactory way. This incompleteness, perhaps, justifies why the problem of how to characterize and control the matter away --- especially very far away --- from the equilibrium is one of the interrelated grand challenges in the U.S. Department of Energy list of roadblocks to progress and the opportunities for truly transformational new understanding \cite{DOE07}. The latter incentive combined with the relentless motivation to unravel any phenomenon in Nature in terms of a single quantum mechanical equation explains a rapid growth in the research focused on the problem of thermalization at a quantum level. 

In fact, statistical mechanics is not expected to work in a quantum system with small number of degrees of freedom or low lying energy states. In these cases, the results obtained are sensitive to the details of the initial {\it pure} state. However, it is generally believed that as the number of degrees of freedom increases, the situation greatly improves and statistical ensemble eventually emerges. From this perspective field theory represents a unique theoretical setup to test the physics of thermalization. 

Remarkably, there are ways to partially elucidate the mechanism of relaxation if a number of plausible assumptions are introduced. One of the prominent frameworks towards understanding how the macroscopic body approaches equilibrium with its environment is the so-called `eigenstate thermalization hypothesis' (ETH) \cite{Jensen85,Deutsch91,Srednicki94}. For a comprehensive review on this subject, the reader is invited to study \cite{Deutsch18,DAlessio:2016rwt}. While the ETH received a lot of attention in the context of many body systems \cite{Gogolin:2016hwy}, a generic field theory setup is still too complex to be addressed by analytic methods, whereas tractable models are truly rare, \eg see \cite{Rigol08,Lashkari:2016vgj} and references therein.

To identify and quantify the universal characteristics of thermalization process it is necessary to analytically control the entire dynamics starting from the very early times, when the system departs from a generic initial state, and ending at late times when the state approaches thermal equilibrium. While the calculations in this paper primarily deal with the early time evolution, we sketch a possible analytic approach to late time dynamics under the assumption that the system obeys thermalization hypothesis. Moreover, to generate a wide class of initial states we consider a unitary process during which a quantum field theory model in its ground state is subject to a generic pulse-like change in the strength of the interaction
\be
 \lambda(t)=\delta\lambda\,f\left(t/\delta t\right)~,
 \label{quenprot}
\ee
where $f(t)$ is a dimensionless smooth function which vanishes sufficiently fast outside some interval of order one centered around $t = 0$. 

The above protocol is conventionally referred to as a global quantum quench, whereas the characteristic inverse time scale, $\delta t^{-1}$, over which the coupling experiences a significant change is called the quench rate. This rate is used to distinguish between fast and adiabatic regimes. 

The interest in quantum quenches received a significant boost due to the successful experiments with cold atoms \cite{Greiner:2002,Bloch:2007,Polkovnikov:2010,Cazalilla:2011,Mitra:2017} where many of the theoretical proposals of quantum thermalization can be tested. This setup exhibits relaxation at late times and is somewhat easier to study experimentally than its counterpart in the field theory regime. 

However, when the the cold atoms system exhibits a quantum critical phase, its dynamics is well approximated by a conformal field theory (CFT) \cite{Kibble:1976sj,Zurek:1985qw}. In particular, one can drive the system through a critical point by changing its optical lattice spacing, while preserving the quantum coherence for a long period of time. When the system approaches criticality the quench rate dominates over all physical scales, adiabatic approximation breaks down, and the entire system is driven far away from the equilibrium. Therefore such systems serve as an ideal setup for experimental and theoretical research where the deep questions, \eg the characteristic time it takes for a system to reach the equilibrium state \cite{Polkovnikov:2010,Cazalilla:2011,Fagotti2013,Mitra:2017}, can be addressed.

It is very difficult to scrutinize a time-dependent process in the field theory context. Even the weak coupling assumption does not help to avoid a serious analytical complexity or shape a way of making late time behavior understandable.  However, for a special class of strongly coupled holographic field theories, the time-dependent processes can be addressed in terms of a gravitational dual \cite{Chesler:2008hg,Basu:2011ft}.  

Quite often holography facilitates the analytic treatment and allows to reveal delicate traits of the dynamics which are either intangible or elusive  within the other approaches.  Thus, for instance, in \cite{Buchel:2012gw,Buchel:2013lla} the authors used  numerical methods in the holographic setup to discern a novel universal scaling with respect to the quench rate in the response of a strongly-coupled CFT to a rapid quench of the scalar and fermionic masses. This result has been generalized and  explained analytically in \cite{Buchel:2013gba}, concluding that at early times the fast quench of a strongly-coupled CFT in $d$ dimensions results in a universal scaling of the observables, \eg $\langle\mathcal{O}\rangle\sim \delta\lambda\delta t^{d-2\Delta}$, where ${d/2}<\Delta < d$ is the scaling dimension of the scalar operator $\mathcal{O}$ adjoint to the quenched coupling $\lambda(t)$.

The universal scaling behavior has further been shown to exist in the context of free field theories \cite{Das:2014jna,Das:2014hqa,Das:2015jka}, and it was argued that the same scaling emerges in any quantum field theory with a UV fixed point provided that the quench rate is sufficiently fast \cite{Das:2016lla,Das:2016eao,Das:2017sgp,Dymarsky:2017awt}. Finally, \cite{Dymarsky:2017awt,Goykhman:2018ihr} used the conformal perturbation theory technique to generalize and extend these results. In particular, the universal scaling behavior of the correlation functions of spinless, spin-${1\over 2}$ and spin-$1$ primaries  was derived.

A smooth quench protocol, characterized by a finite-valued quench rate $\delta t^{-1}$,
can be contrasted with the sudden quench protocol, in which the coupling constant is changed 
instantaneously.
A salient example of the latter is given by the system prepared in the ground state of the 
Hamiltonian $H_\lambda=H_\mt{CFT}+\lambda(t)\,{\cal O}$, where 
$\lambda(t)=\lambda_0\,\theta(-t)$.
Such quenches have been extensively studied in the two-dimensional critical systems,
starting from the earlier works of \cite{Calabrese:2006rx,Calabrese:2007rg}. 
It has been pointed out that the limit $\delta t\rightarrow 0$ of the fast quench regime does not necessarily
match the results of the sudden quench approximation, due to the non-commutativity of the instantaneous  quench
limit and the limit of sending the UV cutoff to infinity first \cite{Buchel:2013lla,Buchel:2013gba,Das:2014jna,Das:2014hqa,Das:2015jka,Das:2016eao,Das:2016lla,Das:2017sgp}.
This is manifested, for instance, in the above scaling relation $\langle\mathcal{O}\rangle\sim \delta\lambda\delta t^{d-2\Delta}$. Obviously, it exhibits divergence in the limit $\delta t\rightarrow 0$ for ${d/2}<\Delta < d$, whereas sudden quench approximation is smooth across  the instant of quench \cite{Calabrese:2006rx,Calabrese:2007rg}.

In the current work we continue to explore the smooth quench protocol \reef{quenprot} in a number of interacting field theory models which are not necessarily scale invariant. We consider one-loop effects and associated RG flow impact on various observables in the field theory models undergoing a quantum quench. Thus, for instance, in section \ref{phi4_section} we elaborate on the mechanism of RG flow that eventually results in a transmutation of the initial quench protocol defined at a certain RG scale into a different protocol at other scales.  

Furthermore, it has been argued in \cite{Das:2016lla,Das:2016eao,Das:2017sgp,Dymarsky:2017awt} that if the quench rate is the shortest scale compared to any other physical scale inherent to the system, then dynamics at early times is dominated by the UV fixed point. Thus it is tempting to employ the conformal perturbation approach to recover leading order effects associated with rapid but smooth quenches. However, in section \ref{3Dmod} we present examples  of the observables where the leading order effects cannot be recovered using the finite order conformal perturbation technique.  An attempt to reproduce these results within the framework of conformal perturbation theory is doomed to fail at any finite order because the perturbative expansion is plagued by IR divergences.

The rest of the paper is organized as follows. In section \ref{check} we provide a simple check of our findings using the ideas of effective action combined with certain results already published in the literature. Besides gaining confidence in the validity of our calculations, this check offers a broader insight into a deep aspect of late time dynamics. The latter is explained and discussed in section \ref{discussion}. We end with Appendix \ref{counterterms}  where the new counterterms which are necessary for two-loop calculations are evaluated.

\section{The indispensable $\phi^4$ theory}
\label{phi4_section}

In this section we explore dynamics triggered by a global quantum quench in the canonical $\phi^4$ model. In $4-\epsilon$ dimensions this model is free in the UV, whereas in the IR limit it is governed by the weakly interacting Wilson-Fisher fixed point. In what follows we explore the global quench \reef{quenprot} of a slightly relevant quartic coupling focusing on the linear response of $\langle\phi^2\rangle$ at the one-loop level. 

It was argued in \cite{Das:2016lla,Das:2016eao,Das:2017sgp,Dymarsky:2017awt} that if the quench rate is large compared to the other physical scales, then evolution of the system is dominated by the Gaussian fixed point. In particular, the linear response of $\langle\phi^2\rangle$ vanishes unless we take into account corrections associated with other relevant couplings of the model.\footnote{The linear response vanishes because $\phi^2$ and $\phi^4$ have different scaling dimensions \cite{Dymarsky:2017awt}.}  Hence, our goal is to evaluate the leading order effect associated with mass of the scalar field by carrying out renormalization of the composite $\phi^2$-operator \cite{Brown:1980qq}. This procedure enjoys a number of advantages over the conformal perturbation technique. It does not suffer from the IR divergences which eventually may lead to a failure of the finite order conformal perturbation theory,\footnote{In the next section we elaborate an example where the finite order conformal perturbation theory breaks down.} and it allows tracking the effects of renormalization group flow on the quench protocol.  Indeed, as we illustrate below, RG flow of the quench protocol is already evident at the one-loop level.

\subsection{Setup and preliminary calculations}

Let us assume that the system at scale $\tilde\mu$ is governed by the following quenched Hamiltonian in $d=4-\epsilon$ dimensions 
\be
 H(t)=\int d^{d-1} \vec x \({1\over 2} \pi^2 + {1\over 2} (\nabla\phi)^2 + {\tilde g_2\over 2}  \phi^2 + {\tilde g_4 \tilde\mu^{4-d}\over 4!} \phi^4 + {\tilde\mu^{4-d}\over 4!} \lambda(t) \phi^4 \)
 \label{Ham}
\ee
We deliberately split the dimensionless coupling constant $\tilde g_4$ from the time-dependent dimensionless profile $\lambda(t)$. We treat them as two types of different vertices. The bare coupling constants (and the bare fields) will be denoted by the (sub)superscript $``0"$. In particular, to linear order in $\lambda(t)$, the bare Hamiltonian takes the form
\be
 H_0(t)=\int d^{d-1} \vec x \({1\over 2} \pi^2_0 + {1\over 2} (\nabla\phi_0)^2 + {g_2^0\over 2}  \phi_0^2 + {g_4^0\over 4!} \phi_0^4 + {c_2^0\over 2} \lambda(t) \phi_0^2 + {c_4^0\over 4!} \lambda(t) \phi_0^4 \) ~.
\ee
In \reef{Ham} we simply assumed that at certain scale of interest, $\tilde\mu$, the renormalized $\tilde c_4$ equals 1, whereas the coupling constants to other possible operators originating from the time-dependent profile $\lambda(t)$
 vanish, \eg $\tilde c_2=0$.\footnote{The list of such possible bare operators (relevant or marginal) includes, but not limited to $\lambda^2(t)\phi^2_0$, $\lambda^2(t)\phi^4_0$, $\lambda^3(t)\phi^2_0$, etc. }  In other words, eq. \reef{Ham} is the definition of our quench protocol. Moreover, while at the scale $\tilde\mu$ the defined quench protocol represents quench of the quartic coupling, it will look differently at other scales. For instance, we will see that due to the RG flow it induces quench of the running mass.  

Using the standard Keldysh-Schwinger formalism one can readily write the linear response of $\phi^2$ to the quench protocol \reef{Ham} at an arbitrary scale $\mu$
\bea
  \langle \phi^2(t,\vec x) \rangle &=&\langle 0|\phi^2|0\rangle + {- i \mu^{4-d} c_4 \over 4!} \int_{-\infty}^t dt' \, \lambda(t') \int d^{d-1} \vec y \, \langle 0| \[\phi^2(t,\vec x), \phi^4(t',\vec y)\] |0\rangle
   \non
   &+& {-i \mu^2 c_2 \over 2} \int_{-\infty}^t dt' \, \lambda(t') \int d^{d-1} \vec y \, \langle 0| \[\phi^2(t,\vec x), \phi^2(t',\vec y)\] |0\rangle
    + \mathcal{O}\(\lambda^2\) ~.
    \label{phi^2}
\eea
Note that so far we account for all orders in $g_2$ and $g_4$. Furthermore, while our quench protocol \reef{Ham} satisfies $\tilde c_2=0$, this coupling flows under the RG flow to a non-zero value at a different scale
$\mu$. In particular, the last term in the above expression represents contribution of $c_2$ at the scale $\mu\neq \tilde\mu$.  

To zeroth order in the coupling constants $g_2$ and $g_4$ the linear response vanishes. Indeed, in this case the second term in the r.h.s. of \reef{phi^2} equals zero because $\phi^2$ and $\phi^4$ have different scaling dimensions, and  their correlation function vanishes identically in a (free) CFT, whereas the last term vanishes because our quench protocol \reef{Ham} does not induce $c_2(\mu)$ in the absence of $g_2$ and $g_4$ (see \reef{ct} below).

Recall now that in $d=4-\epsilon$ dimensions the UV theory is governed by the Gaussian fixed point with $g_4\ll 1$. Thus, in the limit of rapid quenches it makes sense to expand perturbatively in the quartic coupling constant. The leading order result represents a linear response to a particular quench of the free massive scalar field theory. It takes the form 
\bea
  \langle \phi^2(t,\vec x) \rangle &=& \langle 0|\phi^2|0\rangle - { i  \over 2}\Big( {\langle 0|\phi^2|0\rangle\over 2} \mu^{4-d} c_4 +\mu^2 c_2 \Big)
  \non
   &\times& \int_{-\infty}^t dt' \, \lambda(t') \int d^{d-1} \vec y \, 
 \langle 0| \[\phi^2(t,\vec x), \phi^2(t',\vec y)\] |0\rangle\Big|_{g_4=0}
  + \mathcal{O}\(\lambda^2, g_4\lambda, g_4^2\) ~,
  \label{phi2}
\eea
with
\bea
 && \langle 0|\phi^2|0\rangle = {\Gamma\({2-d\over 2}\)\over (4 \pi)^{d/2} } ~ (g_2)^{d-2\over 2}  
 + g_4\mu^{4-d} (g_2)^{d-3} {\Gamma^2\({4-d\over2}\) \over (d-2) (4\pi)^d} +  \mathcal{O}\(g_4^2\) ~, 
  \label{loop}
 \\
&& \langle 0| \[\phi^2(t,\vec x), \phi^2(0)\] |0\rangle\Big|_{g_4=0}= {4 i \, \text{sign}(t) \Theta(-s^2)\over (2\pi)^d} \,\text{Im}\[  e^{-i{\pi(d-2)\over 2}}\({g_2 \over -s^2}\)^{d-2\over 2} K_{d-2\over 2}^2 \(i\sqrt{ -s^2\, g_2}\)\] ~.
 \nonumber
\eea
where $s^2=-t^2 + \vec x^2$ represents interval between the insertion points. For rapid quenches, $g_2 \, \delta t^{\,2}\ll 1  $, one can use the approximation
\bea
  &&\langle 0| \[\phi^2(t,\vec x), \phi^2(0)\] |0\rangle= -i\, {\Gamma\({d-2\over 2}\)^2  \over 4\pi^d}  {\sin(d\pi) \over \(-s^2\)^{d-2}} \, \Theta(-s^2) \, \text{sign}(t)
  \non
  &&
  \times\Big(1 +{-s^2 g_2\over d-4 } + {d-5\over 2(d-6)(d-4)^2} \, (-s^2 g_2)^2 + \mathcal{O}(s^6 g_2^3)\Big)
  \non
  &&+{2 i\over (2\pi)^{d-1} (d-2)} \({g_2\over -s^2}\)^{d-2\over 2}\, \Theta(-s^2) \, \text{sign}(t) \(1 + 2\, { -s^2 g_2 \over d(d-4)} + \mathcal{O}(-s^4 g_2^2) \) ~.
  \label{comm}
\eea
Note that the leading order does not vanish for integer $d$. Instead, it takes the form
\be
   \langle 0| \[\phi^2(t,\vec x), \phi^2(0)\] |0\rangle\Big|_{d\in \mathbb{Z}_+}= {i\, (-1)^{d-2} \over 4\pi^{d-1} } {\Gamma \({d-2\over 2} \)^2 \over \Gamma(d-2)} 
   ~\delta^{(d-3)} (s^2)\, \text{sign}(t) 
   +\ldots~,
\ee
where $d-3$ derivatives of the delta function are taken with respect to its argument. However, we will use the former expression since $d$ is not necessarily integer in our analysis.

Substituting \reef{comm} into \reef{phi2} and integrating over $\vec y$ results in the following linear response function
\bea
  \langle \phi^2(t,\vec x) \rangle &=& {\Gamma\({2-d\over 2}\)\over (4 \pi)^{d/2} } ~ (g_2)^{d-2\over 2} 
  + g_4\mu^{4-d} (g_2)^{d-3} {\Gamma^2\({4-d\over2}\) \over (d-2) (4\pi)^d} 
  \non
  &+& {4\, \cos^2\({d\pi\over 2}\) \Gamma(2-d)\Gamma(d-3)   \over (4\pi)^d}\,(g_2)^{d-2\over 2} \mu^{4-d} c_4 \int_{-\infty}^t dt' {\lambda(t')\over (t-t')^{d-3}}
   \Big(1+  \mathcal{O}\(g_2\delta t^{2}\)  \Big)
  \non
  &+& {64\, \cos^2\({d\pi\over 2}\) \Gamma\({3-d\over 2}\)\Gamma(d-3)   \over (16\pi)^{d+1\over 2}} \mu^2 c_2 \int_{-\infty}^t dt' {\lambda(t')\over (t-t')^{d-3}}
  \Big(1+  \mathcal{O}\(g_2\delta t^{2}\)  \Big)
  \non
   &+& \mathcal{O}\(\lambda^2, g_4\lambda, g_4^2\)\,.
  \non
\eea
The integral over $t'$ experiences a logarithmic divergence in $d=4$. In dimensional regularization this divergence corresponds to a pole which becomes manifest if we rewrite the integral as follows
\bea
  \langle \phi^2(t,\vec x) \rangle &=& {\Gamma\({2-d\over 2}\)\over (4 \pi)^{d/2} } \,(g_2)^{d-2\over 2} 
  + g_4\mu^{4-d} (g_2)^{d-3} {\Gamma^2\({4-d\over2}\) \over (d-2) (4\pi)^d} 
  \non
  &-& {4\, \cos^2\({d\pi\over 2}\) 
  \Gamma(2-d)\Gamma(d-3)   \over (4\pi)^d (d-4)}\,(g_2)^{d-2\over 2} \mu^{4-d} c_4 \int_{-\infty}^t dt' {\Dot \lambda(t')\over (t-t')^{d-4}} 
  \non
  &-& {64\, \cos^2\({d\pi\over 2}\) \Gamma\({3-d\over 2}\)\Gamma(d-3)   \over (16\pi)^{d+1\over 2}(d-4)} \mu^2 c_2 \int_{-\infty}^t dt' {\Dot\lambda(t')\over (t-t')^{d-4}}
  + \ldots   ~,
  \non
  \label{phi22}
\eea
where dot over $\lambda$ denotes derivative with respect to time.

While it seems like we accounted for all the contributions in the calculation of expectation value of $\phi^2$ to linear order in $g_4$ and $\lambda(t)$, this is in fact a fallacy. The reason is two-fold. First, we did not take into account the ordinary counterterms which are necessary to render the correlation functions of $\phi$ finite. Second, we have to renormalize the composite operator $\phi^2$. These calculations are carried out in the next subsection.

\subsection{Counterterms and renormalization of composite operator}

The divergent tadpole diagram contributes linearly (in $g_4$ and $\lambda$) to the full propagator of $\phi$. Hence, the following counterterm must be added to the Lorentzian action to ensure the two-point function is finite\footnote{See Appendix \ref{counterterms}.}
\be
 \delta S_\text{c.t.}={g_2\over 2 (4\pi)^2 (d-4)}  \int d^{d-1}\vec x \int dt \, \big(g_4 + c_4\, \lambda(t) \big) \,\phi^2(t, \vec x) ~.
\ee
As usual, this counterterm combined with the running couplings represent the bare parameters of the theory. To linear order, we thus have 
\bea
g_4^0&=& \mu^{4-d} g_4  ~,\label{ct_0}
 \\
 g_2^0 &=& g_2\Big(1 - {g_4 \over  (4\pi)^2(d-4)}\Big) ~,
 \label{baremass}
  \\
 c_4^0&=& \mu^{4-d} c_4 ~,
 \\
 c_2^0&=&c_2\mu^2 - {g_2 \, c_4 \over  (4\pi)^2(d-4)} ~.
  \label{ct}
\eea
The first two equations are the typical relations that we encounter in the renormalization of the time-independent $\phi^4$-theory. In contrast, the counterterm \reef{ct} is associated with explicit time dependence inherent to the setup. Emergence of such a counterterm is not surprising since time-dependence in general induces new type of vertices under the RG flow. As expected, it is local\footnote{It changes in time, but depends on the instantaneous value of $\lambda(t)$. In fact, there are also counterterms associated with the cosmological constant renormalization, however they drop out from the calculations in the in-in formalism.} and up to time dependence of $\lambda(t)$ it is reminiscent of the mass renormalization \reef{baremass}. However, not all the induced vertices in the effective action are obtained by naive replacement of the coupling constant $g_4$ in the static counterterms with its time-dependent counterpart $c_4\lambda(t)$. We elaborate on some examples of such manifestly new counterterms in Appendix \ref{counterterms}.

Equations  \reef{baremass} and \reef{ct} are the only modifications of the action that have to be accounted for in the calculation of the expectation value of the composite operator $ \phi^2$ to linear order in $g_4$ and $\lambda(t)$.  The contribution of $\lambda(t)$-term associated with the counterterm \reef{ct} is given by
\bea
  \delta_\text{c.t.}\langle \phi^2(t,\vec x) \rangle &=&  { i g_2\, c_4\over 2 (4\pi)^2 (d-4)} \int_{-\infty}^t dt' \, \lambda(t') \int d^{d-1} \vec y \, 
 \langle 0| \[\phi^2(t,\vec x), \phi^2(t',\vec y)\] |0\rangle\Big|_{g_4=0}
 \non
 &=&   { - g_2\, c_4 \,\cos\({d\pi\over 2}\) \, \Gamma\({d-2\over 2}\)\over 8\pi^2  (4\pi)^{d\over 2}(d-3)  (d-4)^2} \int_{-\infty}^t dt'\, {\Dot \lambda(t')\over (t-t')^{d-4}}~.
\eea
Combining with \reef{phi22} gives
\bea
  &&\langle \phi^2(t,\vec x) \rangle = {\Gamma\({2-d\over 2}\)\over (4 \pi)^{d/2} } \,g_2^{d-2\over 2} 
  \(1-{g_4\over 2(4\pi)^2}{d-2\over d-4} - g_4\({m^2\over \mu^2}\)^{d-4\over 2}{\Gamma\({4-d\over 2}\)\over 2(4\pi)^{d/2}}\)
   \\
  &&
  +\bigg( {4\, \cos\({d\pi\over 2}\) \Gamma(5-d)\Gamma(d-3)   \over (4\pi)^{d\over 2} (d-2)}\,\({g_2\over \mu^2}\)^{d-4\over 2} 
  - {  \Gamma\({d-2\over 2}\)\over 8\pi^2} \bigg) 
 {g_2\, c_4 \cos\({d\pi\over 2}\)\over (4\pi)^{d\over 2} (d-3) (d-4)^2} 
 \non
 &&\times \int_{-\infty}^t dt' {\Dot \lambda(t')\over (t-t')^{d-4}} 
 - {64\, \cos^2\({d\pi\over 2}\) \Gamma\({3-d\over 2}\)\Gamma(d-3)   \over (16\pi)^{d+1\over 2}(d-4)} \mu^2 c_2 \int_{-\infty}^t dt' {\Dot\lambda(t')\over (t-t')^{d-4}}
 + \ldots   ~,
 \nonumber
\eea
where we also accounted for the standard mass renormalization \reef{baremass}. The double poles at $d=4$ cancel whereas the simple pole remains
\be
  \langle \phi^2(t,\vec x) \rangle =  {1\over 8\pi^2(d-4)} \, \Bigg[ g_2 + \mu^2 c_2\lambda(t) +  {g_2\,\big(g_4+c_4 \lambda(t)\big) \big(-1+\gamma + \log{g_2\over 4\pi\mu^2}\big) \over 32\pi^2} \Bigg]
   + \text{finite terms}   ~.
  \nonumber
\ee
This pole is removed by renormalizing the bare composite operator $\phi^2$. 

By definition, the renormalized $[\phi^2]$ is given by the mixing of bare $\phi^2$ and identity operator, $\mathbb{I}$. The coefficients are analytic functions of the couplings
\bea
 [\phi^2] &=& Z_{\phi^2} (\lambda,g_4; d) \, \phi^2 + \mu^{d-4} \(g_2 \, Z_0(\lambda,g_4; d) + \Dot\lambda^2(t) \, Z_1(\lambda,g_4; d)
 + \Ddot\lambda(t)\, \, Z_2(\lambda,g_4; d)\) \, \mathbb{I} 
 \non
 &+& \mu^{d-2} c_2 \, Z_c(\lambda,g_4; d) \, \mathbb{I}  ~.
\eea
With the minimal subtraction scheme which we employ, $Z_{\phi^2}$ equals unity plus an ascending series of poles while $Z_0$, $Z_1$, $Z_2$ and $Z_c$ only contain poles in $(d-4)$. To linear order in $g_4$ and $\lambda(t)$, we have
\be
 [\phi^2] = \(1- { g_4+c_4\lambda(t) \over (4\pi)^2(d-4)}+ \ldots\) \, \phi^2 - { 2\, \mu^{d-4}  \over (4\pi)^2(d-4)} \(g_2 + \mu^2 c_2\lambda(t)  - {  g_2\big(g_4+c_4\lambda(t)\big) \over (4\pi)^2(d-4)} + \ldots \) ~.
\ee
where the residues of the poles were fixed by demanding finite $\langle [\phi^2]\rangle$. This relation between the bare and renormalized $\phi^2$ operators is reminiscent of the similar relation in the Euclidean case \cite{Brown:1980qq}. The latter meets our expectations because the manifestly new type of time-dependent counterterms proportional to the temporal derivatives of $\lambda$ emerge at two-loop level only, see Appendix \ref{counterterms}. 

Combining altogether, yields
\bea
  \langle [\phi^2] \rangle &=& {g_2\over 16\pi^2} \, \(\gamma-1 +\log\Big({g_2\over 4\pi\mu^2}\Big)\)  \[ 1 - {c_4\lambda(t)\over 32 \pi^2} 
  \Big(2+\gamma + \log\big(4\pi\mu^2\delta t^2\big) \Big)  \right.
  \\
  && +{g_4\over 32\pi^2}\(\gamma+\log\Big({g_2\over 4\pi\mu^2}\Big)\) 
  -  {c_4\over 16\pi^2} \int_{-\infty}^{t/\delta t} d\xi ~ \log\Big({t\over\delta t} - \xi\Big) {d \lambda(\xi)\over d\xi} \bigg] 
  \non
  &&- {\mu^2 c_2\lambda(t)\over 16\pi^2}\Big(2+\gamma + \log\big(4\pi\mu^2\delta t^2\big) \Big)  
  - {\mu^2\,c_2\over 8\pi^2} \int_{-\infty}^{t/\delta t} d\xi ~ \log\Big({t\over\delta t} - \xi\Big) {d \lambda(\xi)\over d\xi}
  + \ldots ~,
  \nonumber
\eea
where $\xi=t'/\delta t$. The linear response to the quench protocol \reef{Ham} is thus given by
\bea
  &&\delta\langle [\phi^2] \rangle = - {\lambda(t)\over (4\pi)^2} \Big[2+\gamma + \log\big(4\pi\mu^2\delta t^2\big) \Big]
  \Big[\mu^2\,c_2+  {g_2\, c_4\over 32\pi^2} \, \Big(\gamma-1 +\log\big({g_2\over 4\pi\mu^2}\big)\Big)  \Big]
  \non
  && - {1\over 8\pi^2} \int_{-\infty}^{t/\delta t} d\xi ~ \log\Big({t\over\delta t} - \xi\Big) {d \lambda(\xi)\over d\xi}
   \Big[\mu^2\,c_2+ {g_2 \,c_4\over 32\pi^2} \, \Big(\gamma-1 +\log\big({g_2\over 4\pi\mu^2}\big)\Big)  \Big]  +\ldots 
   \label{linres}
\eea
Next we solve the RG equations (\ref{ct_0})-(\ref{ct}) with the boundary conditions, $\tilde c_4=1$ and $\tilde c_2=0$, fixed by the quench protocol \reef{Ham}  
\be
c_4=1+\ldots ~,  \quad c_2={\tilde g_2 \over (4\pi)^2} {\log(\mu/\tilde\mu) \over \mu^2} + \ldots ~,
\ee
where ellipsis encode higher order terms in $g_4$ which contribute to the suppressed corrections in \reef{linres}. As a result, we arrive at
\bea
  \delta\langle [\phi^2] \rangle &=& - {\tilde g_2\over (4\pi)^4} \Big[\gamma-1 +\log\big({\tilde g_2\over 4\pi \tilde \mu^2}\big)  \Big] 
  \times
  \non
  &&\Big[{\lambda(t)\over 2} \(2+\gamma + \log\big(4\pi\mu^2\delta t^2\big)\) + \int_{-\infty}^{t/\delta t} d\xi ~ \log\Big({t\over\delta t} - \xi\Big) {d \lambda(\xi)\over d\xi}\Big]~.
\eea

In fact, the above expression for the linear response is contaminated by the arbitrariness closely related to the choice of counterterms and regularization scheme. Thus, for instance, the presence of the Euler's constant is a clear sign of the minimal subtraction scheme we employ here. The arbitrary constants can be absorbed in the redefinition of the scales $\mu$ and $\tilde\mu$, leaving us with the following universal response function
\be
  \delta\langle [\phi^2] \rangle = - {\tilde g_2\over (4\pi)^4} \log\big({\tilde g_2\over \tilde \mu^2}\big) 
 \Big(\lambda(t) \log(\mu\delta t) + \int_{-\infty}^{t/\delta t} d\xi ~ \log\Big({t\over\delta t} - \xi\Big) {d \lambda(\xi)\over d\xi}\Big)~.
 \label{phi2resp}
\ee

Note that the second term within the parenthesis depends on the shape of the quench profile at all times preceding the instant of observation $t$. It represents a nonlocal history tail. In contrast, the first term is local. It only knows about the instantaneous amplitude of the quench. Of course, the history tail can be ignored relative to the dominating local term in the limit $\delta t\to 0$. However, it prevails at early times after the quench is over, whereas at late times it decays as $\delta t/t$, and one has to resort to higher order calculations.

Furthermore, \reef{phi2resp} is singular in the limit $\delta t\to 0$. This is a salient feature of the fast smooth quenches \cite{Buchel:2013lla,Buchel:2013gba,Das:2014jna,Das:2014hqa,Das:2015jka,Das:2016eao,Das:2016lla,Das:2017sgp,Dymarsky:2017awt,Goykhman:2018ihr}. It shows that this protocol is quite different from an instantaneous quench.

\section{Three-dimensional model}
\label{3Dmod}

In this section we study fast quenches in a three-dimensional model of interacting scalars and Dirac fermions. One loop corrections in this model are finite if dimensional regularization is used, and therefore renormalization is trivial. In particular, calculations become more transparent and easy to follow. 

We set $d=3-\epsilon$ and assume that the system at scale $\mu$ is governed by the following Hamiltonian
\begin{align}
H(t)&=\int d^{d-1}\vec{x}\left(\frac{1}{2}\,\pi^2+\frac{1}{2}(\nabla\phi)^2
+\frac{g_2}{2}\,\phi^2 +\frac{g_4\,\mu^{4-d}}{4!}\,\phi^4 +\frac{g_6\,\mu^{2(3-d)}}{6!}\,\phi^6 \right.
\notag \\
&-i\,\bar\psi\,\vec{\gamma}\cdot\nabla\,\psi +g_2^\psi\,\bar\psi\psi+\left.g_4^\psi\,\mu^{3-d}\,\phi^2\bar\psi\psi
+c\,\lambda(t)\,\mu^{d-\Delta}\,{\cal O}\right)\,,
\label{3d_hamiltonian} 
\end{align}
where $\lambda(t)$ is a dimensionless quench profile, $\Delta$ is the scaling dimension
of ${\cal O}$ and all the couplings are constant in time.

In what follows we study two cases
\begin{align}
{\cal O}_1&=\phi^2\,\bar\psi\psi\,,\quad \Delta_1=2d-3\,,\\
{\cal O}_2&=\frac{1}{6!}\phi^6\,,\quad \Delta_2=3(d-2)\,.
\end{align}
We start from ${\cal O}_1$. The linear response of $\bar\psi\psi$ takes the form
\begin{align}
\delta\langle \bar\psi\psi(t,{\vec x})\rangle &=
-i\,\mu^{3-d}\,c\,\int _{-\infty}^t dt'\,\lambda(t')\int d^{d-1}{\vec y}
\,\langle 0|\left[\bar\psi\psi(t,{\vec x}),
\,\phi(t',{\vec y})^2\,\bar\psi\psi(t',{\vec y})
\right]|0\rangle\notag\\
&=-i\,\langle 0| \phi^2|0\rangle \,\mu^{3-d}\,c\,\int _{-\infty}^t dt'
\,\lambda(t')\int d^{d-1}{\vec y}\,\langle 0|\left[\bar\psi\psi(t,{\vec x}),\,
\bar\psi\psi(t',{\vec y})
\right]|0\rangle\,.\label{deltaO_3d_fermion}
\end{align}

Note that the vacuum expectation value $\langle 0| \phi^2|0\rangle$ is evaluated in the unquenched theory. It is
given by \reef{loop} 
\begin{equation}
\label{phi2loopshort}
\langle 0|\phi^2|0\rangle ={\Gamma\({2-d\over 2}\)\over (4 \pi)^{d/2} }~ (g_2)^{d-2\over 2} 
+{\cal O}(g_4,g_6)\,.
\end{equation} 
Furthermore, the quench profile $\lambda(t)$ cuts the temporal integral in \reef{deltaO_3d_fermion} at a scale of order $\delta t$. Hence, in the fast quench limit   $ g_2^\psi\delta t\ll 1$, we can use the expansion
\begin{equation}
\int d^{d-1}{\vec y}\,\langle 0|\left[\bar\psi\psi(t,{\vec x}),\,\bar\psi\psi(t',{\vec y})\right]|0\rangle
=-i N_\mt{$\bar\psi\psi$} \,\frac{2\pi^\frac{d+1}{2}}{\Gamma\left(\frac{3-d}{2}\right)\,\Gamma(d-1)}\,
\frac{{\rm sgn}\,(t-t')}{|t'-t|^{d-1}}+{\cal O}\big(g_2^\psi |t-t'|\big) ~,
\end{equation}
where $N_\mt{$\bar\psi\psi$}$ is normalization constant defined by
\begin{equation}
\langle 0|\bar\psi\psi(x)\bar\psi\psi(0)|0\rangle =\frac{N_\mt{$\bar\psi\psi$}}{|x|^{2(d-1)}}\,.
\end{equation}

For the Dirac field normalized as in \reef{3d_hamiltonian}, $N_\mt{$\bar\psi\psi$}$ can be fixed using the identity  
\begin{align}
\langle 0|\bar\psi\psi(x)\bar\psi\psi(0)|0\rangle =2^{[d/2]}\,\delta^{\mu\nu}\,
\frac{\partial}{\partial x^\mu}\int \frac{d^dp}{(2\pi)^d}\frac{e^{ip\cdot x}}{p^2}\;
\frac{\partial}{\partial x^\nu}\int \frac{d^dq}{(2\pi)^d}\frac{e^{-iq\cdot x}}{q^2}\,.
\end{align}
Substituting
\begin{equation}
\int \frac{d^dp}{(2\pi)^d}\frac{e^{ip\cdot x}}{p^2}=\frac{\Gamma\left({d-2\over 2}\right)}{4\pi^\frac{d}{2}}\,
\frac{1}{|x|^{d-2}}\,,
\end{equation}
yields
\begin{equation}
\langle 0|\bar\psi\psi(x)\bar\psi\psi(0)|0\rangle =2^{[d/2]}\,\frac{\Gamma\left(\frac{d}{2}\right)^2}{4\pi^d}\,
\frac{1}{|x|^{2(d-1)}}\,.
\end{equation}
Thus,\footnote{
Notice that if $\psi$ is a Majorana fermion then there is an extra contribution to $N_{\bar\psi\psi}$
due to non-vanishing contraction
\begin{equation}
-\langle \bar\psi_\alpha(x)\bar\psi_\beta(0)\rangle \langle \psi^\alpha(x)\psi^\beta(0)\rangle
=\langle \psi^\alpha(x)\bar\psi_\beta(0)\rangle\langle \bar\psi_\alpha(x)\psi^\beta(0)\rangle\,,
\end{equation}
where we used the Majorana condition $\bar\psi _\alpha=\psi^\beta \, C_{\beta\alpha}$.
This gives contribution to $N_{\bar\psi\psi}$ which is equal
to what we have already calculated, and therefore the total value of $N_{\bar\psi\psi}$ will be doubled.}
\begin{equation}
N_\mt{$\bar\psi\psi$}=2^{[d/2]}\,\frac{\Gamma\left(\frac{d}{2}\right)^2}{4\pi^d}\,.
\end{equation}
Combining altogether, we finally arrive at 
\begin{equation}
\delta\langle \bar\psi\psi(t,{\vec x})\rangle = -
\frac{2^{[d/2]-d-1}}{\pi^{d-\frac{1}{2}}}\,\frac{\Gamma\left(1-\frac{d}{2}\right)\,\Gamma\left(\frac{d}{2}\right)^2}
{\Gamma\left(\frac{3-d}{2}\right)\,\Gamma(d-1)}\,(g_2)^{\frac{d}{2}-1}
\,\mu^{3-d}\,c\,\int _{-\infty}^tdt'\,
\frac{\lambda(t')}{(t-t')^{d-1}}\,.
\end{equation}

The constant coefficient in front of the integral vanishes in $d=3$, because one of the gamma functions in the denominator is singular. However, the integral itself diverges at the upper bound $t'=t$, and there is precise balance between the two singularities. To get the $d\to 3$ limit right, we employ integration by parts in the region where the above expression is regular 
\begin{equation}
\delta\langle \bar\psi\psi(t,{\vec x})\rangle = 
\frac{2^{[d/2]-d-3} \, \Gamma\left({d-2\over 2}\right)\,}
{\pi^{d-\frac{3}{2}}\,\Gamma\left(\frac{5-d}{2}\right)\,\Gamma(d-1)\,\sin \left(\frac{\pi\,d}{2}\right)}
(g_2)^{d-2\over 2} \,\mu^{3-d}\,c\,\int _{-\infty}^tdt'\,\frac{\ddot \lambda(t')}{(t-t')^{d-3}}\,.
\end{equation}
Taking now the limit $d\rightarrow 3$, yields
\begin{equation}
\label{psipsiAnswer}
\delta\langle \bar\psi\psi(t,{\vec x})\rangle = -
\frac{\sqrt{g_2}}{32\pi}\,c\,\dot\lambda(t)\,.
\end{equation}
Notice that because of the non-analytic dependence on $g_2$ this expression cannot be obtained by performing a small mass perturbation at any finite order. This can be readily seen at the level of (\ref{phi2loopshort}). In $d=3$ it exhibits a non-analytic dependence on $g_2$, whereas conformal perturbations of $\langle 0|\phi^2|0\rangle=0$ are plagued by infrared divergences.

Consider now ${\cal O}_2$. The linear response of $\phi^4$ takes the form
\begin{align}
\delta\langle\phi(t,{\vec x})^4\rangle &=
-i\,\frac{\mu^{2(3-d)}\,c}{6!}\,\int _{-\infty}^t dt'\,\lambda(t')\int d^{d-1}{\vec y}\,\langle 0|\left[\phi(t,{\vec x})^4,
\,\phi(t',{\vec y})^6
\right]|0\rangle\notag\\
&=-i\,\frac{\mu^{2(3-d)}\,c}{48}
\,\langle 0| \phi^2|0\rangle \,\int _{-\infty}^t dt'\,\lambda(t')\int d^{d-1}{\vec y}\,\langle 0|\left[\phi(t,{\vec x})^4,\,
\phi(t',{\vec y})^4
\right]|0\rangle\,.\label{deltaO_3d_phi6}
\end{align}
In the fast quench limit we can make use of the following expansion
\begin{equation}
\int d^{d-1}{\vec y}\,\langle 0|\left[\phi(t,{\vec x})^4,\,\phi(t',{\vec y})^4\right]|0\rangle
=-iN_{\phi^4}\,\frac{2\pi^\frac{d+1}{2}}{\Gamma\left(\frac{9-3d}{2}\right)\,\Gamma\left(2(d-2)\right)}
\,\frac{{\rm sgn}(t-t')}{|t'-t|^{3d-7}}+\ldots,
\end{equation}
where normalization constant is defined by 
\begin{equation}
\langle \phi(x)^4\,\phi(y)^4\rangle=\frac{N_{\phi^4}}{|x-y|^{4(d-2)}}\,.
\end{equation}
For canonically normalized scalar field, we have
\begin{equation}
\langle \phi(x)^4\,\phi(y)^4\rangle=4!\,\langle \phi(x)\,\phi(y)\rangle^4
=\frac{3\,\Gamma\left(\frac{d-2}{2}\right)^4}{32\pi^{2d}}\,\frac{1}{|x-y|^{4(d-2)}}\,.
\end{equation}
Hence,
\begin{equation}
N_{\phi^4}=\frac{3\Gamma\left(\frac{d-2}{2}\right)^4}{32\pi^{2d}}\,.
\end{equation}
Combining, yields
\begin{equation}
\delta\langle\phi(t,{\vec x})^4\rangle =
-\frac{\Gamma\left(\frac{d-2}{2}\right)^4\,\Gamma\left(1-\frac{d}{2}\right)\,(g_2)^{d-2\over 2}\,\mu^{2(3-d)}\,c}
{2^{d+8}\,\pi^{2d-\frac{1}{2}}\,\Gamma\left(\frac{9-3d}{2}\right)\,\Gamma(2(d-2))}
\,\int_{-\infty}^tdt'\,\frac{\lambda(t')}{(t-t')^{3d-7}}\,.
\end{equation}
Or equivalently,
\begin{equation}
\delta\langle\phi(t,{\vec x})^4\rangle =
\frac{\Gamma\left(\frac{d-2}{2}\right)^4\,\Gamma\left(1-\frac{d}{2}\right)\,(g_2)^{\frac{d}{2}-1}\,
\mu^{2(3-d)}\,c\,\dot\lambda(t)}
{2^{d+9}\,\pi^{2d-\frac{1}{2}} \,\Gamma\left(\frac{11-3d}{2}\right)\,\Gamma(2(d-2))\,(3d-8)}+{\cal O}(d-3)\,.
\end{equation}
In the limit $d\rightarrow 3$ we finally obtain 
\begin{equation}
\delta\langle\phi(t,{\vec x})^4\rangle =-\frac{\sqrt{g_2}}{2^{11}\pi^3}\,c\,\dot\lambda(t)\,.
\label{phi4Ans}
\end{equation}
This expression is similar in structure to (\ref{psipsiAnswer}) because $\bar\psi\psi$ and $\phi^4$ have identical scaling dimensions in $d=3$.

\section{Alternative derivation and interpretation}
\label{check}

In this section we focus on an alternative explanation of the results obtained in the previous sections. We start from the quench protocol \reef{Ham}. It induces a time-dependent quantum correction to the mass of the excitations governed by the unquenched Hamiltonian. The effective mass at one-loop level is given by
\be
 g_2^\mt{eff}=\tilde g_2 + {\tilde g_4 \over 2}\, \tilde\mu^{4-d} \langle0|\phi^2|0\rangle  + {\lambda(t)\over 2}\, \tilde\mu^{4-d} \langle0|\phi^2|0\rangle~,
 \label{effmass}
\ee
where $|0\rangle$ is the vacuum state of the unperturbed theory. Thus, the time-dependent interaction term can be absorbed in the redefinition of mass, and the original quench protocol \reef{Ham} transmutes into the quench of mass in a free field theory. Now it was shown in \cite{Das:2014jna,Das:2014hqa,Dymarsky:2017awt} that rapid quench of the mass parameter in a scalar {\it free} field theory results in the following universal behavior
\be
 \langle[\phi^2]\rangle= - {m^2(t)\over 8\pi^2} \log(\mu\delta t) ~,
\ee
where $m^2(t)$ is the profile of quench. In our case $m^2(t)$ should be identified with the induced quench protocol \reef{effmass}, \ie $m^2(t)={\lambda(t)\over 2}\, \tilde\mu^{4-d} \langle0|\phi^2|0\rangle$. Hence, we expect the following linear response 
\be
 \langle[\phi^2]\rangle= - {\lambda(t)\tilde\mu^{4-d}\over 16 \pi^2} \langle0|\phi^2|0\rangle \log(\mu\delta t) ~.
\ee
Indeed, it matches \reef{phi2resp} upon substituting $d=4$ and\footnote{To get this result, one has to evaluate 
\be
  \langle0|\phi^2|0\rangle=\int_{p<\tilde\mu} {d^4 p\over (2\pi)^4} {1\over p^2 + \tilde g_2} ~,
\ee
and keep the universal terms in the limit $\tilde\mu\gg \tilde g_2$.}
\be
 \langle0|\phi^2|0\rangle={\tilde g_2\over 16\pi^2} \log  \({\tilde g_2\over 4\pi\tilde\mu^2}\) ~.
\ee

Similarly, it was argued in \cite{Das:2014hqa,Dymarsky:2017awt} that fast quench of the mass parameter in the fermionic {\it free} field theory yields
\be
  \langle[\bar\psi\psi]\rangle= {1\over 8} {d m(t)\over dt}
\ee
where this time $m(t)$ represents quench profile of the fermionic mass.

Now, the induced one-loop correction to the effective mass of the fermion in a theory governed by \reef{3d_hamiltonian} reads
\be
 (g_{2}^\psi)^\mt{eff}=g_2^\psi +g_4\mu^{3-d} \langle0|\phi^2|0\rangle + c\lambda(t)\mu^{3-d} \langle0|\phi^2|0\rangle. 
\ee
Hence, $m(t)= c\lambda(t) \mu^{3-d} \langle0|\phi^2|0\rangle$. Substituting $d=3$ and $\langle0|\phi^2|0\rangle= -\sqrt{g_2}/(4\pi)$, gives 
\be
  \langle[\bar\psi\psi]\rangle= - { \sqrt{g_2}\over 32\pi} \, c {d \lambda(t)\over dt} ~,
\ee
in full agreement with \reef{psipsiAnswer}.

Obviously, not all the leading order results can be recovered based on the manipulations with quadratic effective action. For instance, at the one-loop level $\mathcal{O}_2$ induces quench of the quartic coupling in \reef{3d_hamiltonian}
\be
 g_4^\mt{eff}=g_4+{g_6 \over 48} \, \mu^{2(3-d)} \langle0|\phi^2|0\rangle
 +{c\lambda(t) \over 48} \, \mu^{2(3-d)} \langle0|\phi^2|0\rangle~.
\ee
Substituting now the time-dependent quartic coupling $c\lambda(t)\mu^{2(3-d)}\langle0|\phi^2|0\rangle/48$ along with $\Delta=2$ and $d=3$ into (3.7) of \cite{Dymarsky:2017awt} we reproduce \reef{phi4Ans}.

\section{Discussion}
\label{discussion}

The main aspiration of the current work is to advance our understanding of the out of equilibrium dynamics in the quantum field theory context. We continued to explore a new scaling regime originally revealed in a strongly-coupled CFT with holographic dual \cite{Buchel:2012gw,Buchel:2013lla,Buchel:2013gba}. Previous efforts were mostly focused on the early time evolution of the system launched by the quench of either free \cite{Das:2014jna,Das:2014hqa,Das:2015jka,Das:2016lla,Das:2016eao,Das:2017sgp} or interacting field theories in the vicinity of the UV fixed point \cite{Dymarsky:2017awt,Goykhman:2018ihr}. In particular, conformal perturbation technique was employed to foster the analytic studies in the interacting case. In this paper, however, we studied various aspects of quantum quenches which are not accessible within the framework of conformal perturbation theory.

Thus, for instance, we considered an interacting scalar field theory model in $4-\epsilon$ dimensions subject to a pulse-shaped quench of the quartic coupling constant. We conducted renormalization of this model at one-loop level, and showed that RG flow results in a transmutation of the quench protocol defined at a certain energy scale into a different looking protocol at other scales. This phenomenon is generic, and one should expect it in any interacting field theory subject to renormalization.  Here we explicitly tracked this mechanism showing that quench of the quartic coupling eventually induces quench of the running mass.

The scaling expression \reef{phi2resp} for the $\phi^2$ exhibits a non-analytic dependence on the coupling $ g_2$. Naively this suggests that one cannot reproduce \reef{phi2resp} within the framework of conformal perturbation technique. However, this conclusion is  premature.  We provide now a sketch of such a calculation which ultimately results in \reef{phi2resp}. It demonstrates explicitly  the weaknesses of the method. These flaws are manifest in the analysis of the three-dimensional model, where the conformal perturbation technique happens to be inadequate approach to get our results at any finite order in the perturbative expansion.

We start from noting that one has to resort to the second order perturbation theory to recover \reef{phi2resp}. Schematically the relevant term is given by \cite{Dymarsky:2017awt}
 \be
  \langle \phi^2(t,\vec x) \rangle \sim {g_2\over 2}\int dt_1 {\mu^{4-d}\lambda(t_1)\over 4!} \int d^{d-1} {\vec y_1} \int dt_2 
  \int d^{d-1} {\vec y_2}  ~
  \langle0|  \phi^4(t_1,\vec y_1) \phi^2(t, \vec x) \phi^2(t_2,\vec y_2)|0\rangle ~,
  \label{phi2cft}
 \ee
where the equality holds iff the right ordering of operators is imposed within the vev. We ignore it since precise ordering is not essential for the punchline. If necessary it can be restored using the Keldysh-Schwinger formalism.\footnote{See \cite{Dymarsky:2017awt} where the analogous calculation was explicitly carried out.} 

The three point function on the right hand side of \reef{phi2cft} is evaluated in the massless free field theory. Hence, it takes the following simple form
\bea
 &&\langle0|  \phi^4(t_1,\vec y_1) \phi^2(t, \vec x) \phi^2(t_2,\vec y_2)|0\rangle = 
 4! \langle0|  \phi(t_1,\vec y_1) \phi(t, \vec x)|0\rangle^2 \langle0|  \phi(t_1,\vec y_1)  \phi(t_2,\vec y_2)|0\rangle^2
 \\
 &&\quad\quad\quad= 
 {4! ~\Gamma^4\({d-2\over 2}\)\over 16^2 \pi^{2\,d} \Big( (\vec x-\vec y_1)^2 - (t_1-t-i\epsilon)^2\Big)^{d-2}  
 \Big( (\vec y_1-\vec y_2)^2 - (t_1-t_2-i\epsilon)^2\Big)^{d-2} } ~,
 \nonumber
\eea
where the $i\epsilon$ prescription matches ordering of operators in the first line. As a result, the integrals over $t_1, \vec y_1$ and $t_2, \vec y_2$ in \reef{phi2cft} decouple and can be carried out separately. Simple power counting reveals that both of them exhibit logarithmic UV divergence in $d=4$. These divergences are regulated by cutting the integrals at the UV scale $\mu$. This is the origin of the double logarithm in \reef{phi2resp}. In addition, the integral over $t_2, \vec y_2$ suffers from the IR divergence, and it is natural to identify the IR regulator of this integral with the mass parameter $g_2$. Obviously, there are no other relevant scales.\footnote{In contrast, the IR end of the integral over $t_1, \vec y_1$ is regulated by the decaying coupling $\lambda(t)$ since we assume $g_2\delta t^2\ll 1$. This is the origin of the scale $\delta t$ in one of the logarithms in \reef{phi2resp}.} As a result, one ends up with \reef{phi2resp}.

Even though the conformal perturbation approach experienced IR flaws in the above analysis, we were able to reproduce  \reef{phi2resp} because it is  entirely fixed by the UV modes.  However, the finite order perturbative expansion around the fixed point will necessarily fail if the leading order result is sensitive to the IR.  We demonstrated this point by studying fast quenches in the three-dimensional model of interacting scalar and Dirac fields. The leading order linear response functions \reef{psipsiAnswer} and \reef{phi4Ans} are manifestly non-analytic in the mass parameter $g_2$. It is not feasible to get these results at any finite order of the perturbative expansion around zero mass. We pointed out that the loop effects in this model are dominated by the IR modes. In particular, perturbative expansion around $g_2=0$ is ill-defined at any finite order because of IR divergences. Thus conformal perturbation series must be re-summed.  

Furthermore, time-dependent loop corrections to the effective action can be regarded as a particular quench protocol of its time-independent sector. Our calculations fit consistently into this framework. In the simplest case, quench of the coupling constant is effectively described as  quench of the mass parameter in a free theory. We confirmed this interpretation using the canonical $\phi^4$ field theory and the model of interacting fermions and scalars in three dimensions. We used \cite{Das:2014jna,Das:2016lla,Dymarsky:2017awt} to argue that \reef{phi2resp} and \reef{psipsiAnswer} match the linear response functions induced by the rapid change in the effective scalar and fermionic masses respectively. Moreover, this interpretation can be also extended to effective interactions. Specifically we used \cite{Dymarsky:2017awt} to demonstrate how \reef{phi4Ans} emerges from the linear response generated by the fast quench of the effective quartic coupling.

The effective action approach in the context of quenches might be useful to understand better the process of thermalization. Indeed, it was argued in \cite{Dymarsky:2017awt} that if the quench rate is sufficiently fast, then all local quantities equilibrate to their thermal values specified by an excess energy acquired by the system during the quench. In particular, the effective mass at late times should play a role of the thermal mass if the state of the system approaches thermal ensemble in the above sense. Hence, one can estimate the effective temperature by identifying late time effective mass with its thermal counterpart. 

Of course, evaluating the effective mass in a time-dependent setup sounds like a horrendous challenge. The linear response analysis presented in this work will either decay or explode as $t\to\infty$ indicating that the late time dynamics is entirely dominated by a non-linear regime which is not tractable within the perturbative approach. However, for a certain class of quenches this calculation is analytically approachable. 

Indeed, it was pointed out in \cite{Das:2014jna} that mass quenches in a free field theory are equivalent to evolution in a cosmological background  provided that the time-dependent mass follows a smooth hyperbolic profile. In particular, exact analytic solution studied previously in the cosmological context \cite{Bernard:1977pq,Birrell:1982ix} was extended and used to explore various aspects of quantum quenches \cite{Das:2014jna,Das:2014hqa,Das:2015jka,Das:2016eao,Das:2016lla,Das:2017sgp}. We propose to endow this setup with constant interaction and use full propagator that encodes all orders in the quench parameter to evaluate effective mass perturbatively in the coupling {\it constant}. This suggests an interesting avenue for further research and we plan to pursue it in the future.

It was argued recently \cite{Alves:2018qfv,Camargo:2018eof} that complexity can serve as a novel probe of quantum quenches. It would be interesting to extend the proposed definitions \cite{Jefferson:2017sdb,Chapman:2017rqy} beyond the free field theory context to understand if complexity shapes a powerful tool towards unravelling the conundrums of thermalization in the fields theory context, see \cite{Bhattacharyya:2018bbv} for recent progress in this direction.

\acknowledgments  
We thank Soumangsu Chakraborty, Diptarka Das, Anatoly Dymarsky, Shmuel Elitzur and Ruth Shir for helpful discussions and comments.  This work is supported by the Binational Science Foundation (grant No 2016186) and by the "Quantum Universe" I-CORE program of the Israel Planning and Budgeting Committee (grant No. 1937/12).

\appendix

\section{Counterterms}
\label{counterterms}

Let us introduce a background field $\bar\phi$ in the path integral representation of $|\Psi(t)\rangle$. To second order in $g_4^r(t)\equiv g_4+c_4 \lambda(t)$ we get the following {\it quadratic} in $\bar\phi$ contribution to the effective action 
\bea
 &&i\delta\Gamma_\text{eff}(\bar\phi)=\({-i\mu^{4-d}\over 4}\) \int_{-\infty}^t dt' g_4^r(t') \int d^{d-1}\vec x ~ \bar\phi^2(t',\vec x) \, \langle 0| \phi^2|0\rangle
 +{\mu^{2(4-d)} \over 4} \({-i\over 2}\)^2 \, \langle 0| \phi^2 |0\rangle
 \non
 &&~ \times ~ \int_{-\infty}^t dt' g_4^r(t') \int d\vec x \, \bar\phi^2(t',\vec x) \int_{-\infty}^t dt'' g_4^r(t'') \int d\vec y 
 \langle 0| \, \mathcal{T}\Big( \phi^2(t',\vec x) ~ \phi^2(t'', \vec y)\Big) |0\rangle 
 \non
 && ~ + ~{\mu^{2(4-d)} \over 2} \({-i\over 3!}\)^2
  \label{effact}
 \\
 && ~ \times \int_{-\infty}^t dt' g_4^r(t') \int d\vec x \, \bar\phi(t',\vec x) \int_{-\infty}^t dt'' g_4^r(t'') \int d\vec y \, \bar\phi(t'',\vec y)
 \langle 0| \, \mathcal{T}\Big( \phi^3(t',\vec x) ~ \phi^3(t'', \vec y)\Big) |0\rangle~,
 \nonumber
\eea
where all correlation functions are implicitly connected. It follows from \reef{loop} that $\langle 0| \phi^2|0\rangle$ diverges in $d=4$, and therefore it is necessary to introduce the counterterms \reef{baremass} and \reef{ct} to ensure finiteness of the effective action to linear order in $g_4^r$. The relation \reef{baremass} is essentially the textbook counterterm encountered in the static case, whereas \reef{ct} is obtained from \reef{baremass} by a mere replacement $g_4\to c_4$. From this perspective these counterterms are cousins. 

Obviously, many of the time-dependent counterterms can be obtained by substituting the static coupling with its time-dependent counterpart. In particular, the second term in \reef{effact} induces a counterterm of this type. However, by far such terms do not exhaust the list of new time-dependent counterterms. Thus, for instance, a new type of time-dependent counterterms shows up in the last term of the above expression, and we elaborate it in what follows.

We start from recalling that divergences in various terms of the perturbative expansion appear when the local operators, \eg two $\phi^3$ operators in the 3rd term of \reef{effact}, approach each other due to integration over the insertion points. Therefore one can employ OPE technique to recover counterterms associated with local divergences. This procedure is simple in the vicinity of the Gaussian theory. First, we expand the background field $\bar\phi(t'',\vec y)$ to second order\footnote{Higher orders do not contribute to divergences in four dimensions. One can also expand the time-dependent coupling $g_4^r(t'')$ around the instant $t'$, however we find it less convenient.} around the insertion point $(t',\vec x)$
\bea
 &&i\delta\Gamma_\text{eff}(\bar\phi)\supset{\mu^{2(4-d)} \over 2} \({-i\over 3!}\)^2  
 \int_{-\infty}^t dt'  \int d\vec x \, g_4^r(t') \, \bar\phi(t',\vec x) \int_{-\infty}^t dt'' g_4^r(t'') 
 \non
 && ~ \times\int d\vec y  \Big(\bar\phi(t',\vec x) + (y-x)^\mu\del_\mu\bar\phi(t',\vec x)+{(y-x)^\mu(y-x)^\nu\over 2} \del_\mu\del_\nu \bar\phi(t',\vec x) + \ldots\Big) D(t',\vec x\,;\, t'',\vec y),
 \nonumber
\eea
where
\bea
D(t',\vec x\,;\, t'',\vec y)&\equiv&\langle 0| \, \mathcal{T}\big( \phi^3(t',\vec x) ~ \phi^3(t'',\vec y)\big) |0\rangle = 3! \langle 0| \, 
\mathcal{T}\big( \phi(t',\vec x) ~ \phi(t'',\vec y)\big) |0\rangle^3
\\
&=& {6\over (2\pi)^{3d\over 2}} \({g_2\over s^2}\)^{3(d-2)\over 4} K_{d-2\over 2}^{\,3}\(\sqrt{g_2 \, s^2}\)~,
\eea
with $s^2=-(t'-t'')^2+(\vec x - \vec y)^2 + i\epsilon$. 

To extract the UV divergences we can expand $D(t',\vec x\,;\, t'',\vec y)$ in the limit of coincident points, $t''\to t'$ and $\vec y\to \vec x$, which is tantamount to OPE argument we mentioned above. Suppressing the higher order terms which do not result in the four dimensional divergent integrals, gives 
\bea
 D(t',\vec x\,;\, t'',\vec y)&=& {3 \,\Gamma^2\({d-2\over 2}\)\over 32(\pi)^{3d\over 2} ( s^2)^{3(d-2)\over 2} } \( \Gamma\Big({d-2\over 2}\Big) 
 \Big( 1 - {3\over 2(d-4)}  \, g_2\, s^2 +\mathcal{O}(g_2^2 s^4) \Big)  \right.
 \non
   &+& \left. {3\over 2^{d-2}}\, \Gamma\Big({2-d\over 2}\Big)  (g_2 s^2)^{d-2\over 2} \(1+\mathcal{O}(g_2 s^2) \)  \) ~.
\eea
The divergent structure of the effective action thus takes the form\footnote{We keep only terms which exhibit poles in $d=4$. The IR divergences should not be taken at face value, they are regulated by finite mass, and we simply ignore them.}
\bea
 &&\delta\Gamma_\text{eff}^\text{div}(\bar\phi)\supset  {i\,\Gamma^2\({d-2\over 2}\) \over 16^2 (\pi)^{3d\over 2} }\,  \mu^{2(4-d)}
 \int_{-\infty}^t dt'  \int d\vec x \, g_4^{r}(t') \, \bar\phi^2(t',\vec x) \int_{-\infty}^t dt'' g_4^{r}(t'')
 \non
 &&  \times   \int d\vec y  {1\over ( s^2)^{3(d-2)\over 2} } 
 \( {\Gamma\({d-2\over 2}\)\over 3 }  + {\Gamma\Big({2-d\over 2}\Big) \over 2^{d-2}}  (g_2 s^2)^{d-2\over 2} - {\Gamma\Big({d-2\over 2}\Big)\over 2(d-4)} \, g_2\, s^2 \)
 \non
&&+  \,   {i\,\Gamma^3\({d-2\over 2}\)\over 3\cdot 16^2 (\pi)^{3d\over 2}  } \mu^{2(4-d)} 
 \int_{-\infty}^t dt'  \int d\vec x \, g_4^r(t') \, \bar\phi(t',\vec x) \del_\mu\bar\phi(t',\vec x) \int_{-\infty}^t dt'' g_4^r(t'') 
 \int d\vec y  \, {(y-x)^\mu\over ( s^2)^{3(d-2)\over 2}}
 \non
 \non
&&  + \, {i\,\Gamma^3\({d-2\over 2}\)\over 6\cdot 16^2(\pi)^{3d\over 2} }   \mu^{2(4-d)}
 \int_{-\infty}^t dt'  \int d\vec x  \, g_4^{r}(t') \bar\phi(t',\vec x) \del_\mu\del_\nu \bar\phi(t',\vec x) 
   \int_{-\infty}^t dt'' g_4^{r }(t'') 
\non
&&\times \int d\vec y  \, { (y-x)^\mu(y-x)^\nu \over ( s^2)^{3(d-2)\over 2} } 
\eea
Next we integrate over $\vec y$
\bea
 &&\delta\Gamma_\text{eff}^\text{div}(\bar\phi)\supset  {i\,\Gamma^2\({d-2\over 2}\) \over 16^2 (\pi)^d \sqrt{\pi} }\,  \mu^{2(4-d)}
 \int_{-\infty}^t dt'  \int d\vec x \, g_4^{r}(t') \, \bar\phi^2(t',\vec x) \int_{-\infty}^t dt'' {g_4^{r}(t'') \over \( -(t'-t'')^2 +i\epsilon\)^{2d-5\over 2}}
 \non
 && ~ \times   
 \( {\Gamma\({d-2\over 2}\) \Gamma\( {2d-5\over 2} \) \over 3 \, \Gamma\big( {3(d-2)\over 2} \big) }  
 + {\Gamma\big({2-d\over 2}\big) \Gamma\({d-3\over 2}\)\over 2^{d-2} \, \Gamma(d-2)}  \( -g_2(t'-t'')^2 +i\epsilon\)^{d-2\over 2} 
 \right.
 \non
 && \quad\quad\quad 
 \left.- {\Gamma\big({d-2\over 2}\big) \Gamma\( {2d-7\over 2} \)\over 2(d-4) \Gamma\big( {3d-8\over 2} \big)} \, 
  \( -g_2(t'-t'')^2 +i\epsilon\)    \)
 \non
 &&+\,
 {i\,\Gamma^3\({d-2\over 2}\) \Gamma\( {2d-5\over 2} \) \over 3\cdot 16^2 (\pi)^d \sqrt{\pi} \Gamma\big( {3(d-2)\over 2} \big) } 
 \mu^{2(4-d)} 
 \non
 &&\times \int_{-\infty}^t dt'  \int d\vec x \, g_4^r(t') \, \bar\phi(t',\vec x) \del_{t'}\bar\phi(t',\vec x) \int_{-\infty}^t dt'' g_4^r(t'') 
  {(t''-t') \over \( -(t'-t'')^2 +i\epsilon\)^{2d-5\over 2} }
 \non
 \non
&&  + {i\,\Gamma^3\({d-2\over 2}\) \Gamma\( {2d-7\over 2} \) \over 12\cdot 16^2(\pi)^d \sqrt{\pi} \Gamma\big( {3(d-2)\over 2} \big)}   \mu^{2(4-d)} \( \delta^\mu_j \delta^\nu_j - (2d-7) \delta^\mu_0 \delta^\nu_0  \)
\non
&&\times  
 \int_{-\infty}^t dt'  \int d\vec x  \, g_4^{r}(t') \bar\phi(t',\vec x) \del_\mu\del_\nu \bar\phi(t',\vec x) 
   \int_{-\infty}^t dt'' {g_4^{r}(t'') \over \( -(t'-t'')^2 +i\epsilon\)^{2d-7\over 2}}
\eea

The temporal integrals over $t''$ exhibit logarithmic divergence in $d=4$. In dimensional regularization they correspond to poles. To get the pole  structure we use the following simple identity
\be
 \int_0^\infty dy \, y^{\epsilon-1} f(y) = {-1\over \epsilon} \int_0^\infty dy \, y^{\epsilon} {d\over dy} f(y) = {f(0)\over \epsilon} 
 - \int_0^\infty dy \, (\log y)  {d\over dy} f(y) ~, \quad 0<\epsilon\ll 1~,
\ee
where $f(y)$ is a smooth function decaying sufficiently fast as $y\to \infty$. As a result, we get\footnote{The signature of Minkowski metric is mostly plus.}
\bea
 &&\delta\Gamma_\text{eff}^\text{div}(\bar\phi)\supset  {- 1 \over 2 (4\pi)^4 }\,  {g_2\over (d-4)^2}\(1 + \mathcal{O}\big( d-4 \big) \)
 \int_{-\infty}^t dt'  \int d\vec x \, g_4^{r 2}(t') \, \bar\phi^2(t',\vec x)
 \non
&&  + \, {-1\over 24 \, (4\pi)^4 \,(d-4)} 
 \int_{-\infty}^t dt'  \int d\vec x  \, g_4^{r2}(t') \bar\phi(t',\vec x) \del^2 \bar\phi(t',\vec x) 
 \non
 &&+\,
 {-1 \over 12  (4\pi)^4 (d-4) }  \int_{-\infty}^t dt'  \int d\vec x \, g_4^r(t')\del_{t'} g_4^r(t') \, \bar\phi(t',\vec x)  \del_{t'}\bar\phi(t',\vec x) 
 \non
&&  + \, {1\over 24 \, (4\pi)^4 \,(d-4)} 
 \int_{-\infty}^t dt'  \int d\vec x \, g_4^{r}(t') \del_{t'}^2g_4^{r}(t')\bar\phi^2(t',\vec x) ~.
\eea
We suppressed a simple pole in the first term. Its residue is contaminated with arbitrariness associated with the precise choice of $\mu$. As usual, this sort of arbitrariness disappears if all the divergences are taken into account.

Note that divergences in the first and second terms survive in the static limit and match textbook calculations, whereas the third and fourth terms are inherent to the time-dependent coupling $g_4^r(t)$. These divergences are eliminated by adding an appropriate set of counterterms to the effective action. In particular, if the quartic coupling depends on time, then it is necessary to supplement the standard set of counterterms with
\bea
 \delta S_\text{c.t.}&=& 
 {1 \over 12  (4\pi)^4 (d-4) }  \int_{-\infty}^t dt'  \int d\vec x \, g_4^r(t')\del_{t'} g_4^r(t') \, \bar\phi(t',\vec x)  \del_{t'}\bar\phi(t',\vec x) 
 \non
 &+& {- 1\over 24 \, (4\pi)^4 \,(d-4)} 
 \int_{-\infty}^t dt'  \int d\vec x \, \, g_4^{r}(t') \del_{t'}^2g_4^{r}(t') \bar\phi^2(t',\vec x) ~. 
 \label{newCT}
\eea

More generally, if $g_4$ is space-time dependent, then these counterterms take the form
\bea
 \delta S_\text{c.t.}&=&
  {-1 \over 12  (4\pi)^4 (d-4) }  \int_{-\infty}^t dt'  \int d\vec x \, g_4^r(t')\del_\mu g_4^r(t') \, \bar\phi(t',\vec x)  \del^\mu\bar\phi(t',\vec x) 
 \non
 &+&  {1\over 24 \, (4\pi)^4 \,(d-4)} 
 \int_{-\infty}^t dt'  \int d\vec x ~ g_4^{r}(t', \vec x) \, \del^2 g_4^{r}(t',\vec x) ~ \bar\phi^2(t',\vec x) ~. 
\eea
This conclusion follows immediately from \reef{newCT} if one thinks of $g_4^r$ as an external Lorentz scalar and imposes Lorentz invariance. 

So far the discussion refers to the path integral representation of $|\Psi(t)\rangle$, or equivalently, the counteterm \reef{newCT} is associated with the upper half of the Keldysh-Schwinger contour. However, similar considerations apply for the path integral representation of $\langle\Psi(t)|$. Hence, in the Keldysh-Schwinger formalism this countertrem is living on the entire closed time contour.

%

\end{document}